# Driving Factors Behind the Social Role of Retail Centers on Recreational Activities

Sepideh Baghaee[1], Saeed Nosratabadi[2], Farshid Aram[1], and Amir Mosavi[3,4,5]*

[1] Escuela Técnica Superior de Arquitectura, Universidad Politécnica de Madrid-UPM, Madrid 28040, Spain; Sepideh.baghaee@alumnos.upm.es, Farshid.aram@alumnos.upm.es

[2] Doctoral School of Economic and Regional Sciences, Hungarian University of Agriculture and Life Sciences, Páter Károly u. 1, 2100, Gödöllő, Hungary; saeed.nosratabadi@phd.uni-szie.hu

[3] Faculty of Civil Engineering, Technische Universität Dresden, 01069 Dresden, Germany

[4] Norwegian University of Life Sciences, School of Economics and Business, 1430 Oslo, Norway

[5] Obuda University, 1034 Budapest, Hungary

*Correspondence: Amir Mosavi: amirhosein.mosavi@nmbu.no; amir.mosavi@mailbox.tu-dresden.de

**Abstract:** Retail centers can be considered as places for interactional and recreational activities and such social roles of retail centers contribute to the popularity of the retail centers. Therefore, the main objective of this study was to identify effective factors encouraging customers to engage with interactional activities and measure how these factors affect customer behavior. Accordingly, two hypotheses were raised illustrating that the travel time (i.e., the time it takes for a customer to reach the retail center) and the variety of shops (in a retail center) increase the percentage of people who spend their leisure time and recreational activities retail centers. Two case studies were conducted in two analogous retail centers, one in Tehran, Iran, and the other in Madrid, Spain. According to the results, there is an interaction between the travel time and the motivation for the presence of people in the retail center. Furthermore, the results revealed that half of both retail center goers who spend more than 10 minutes to reach the retail centers prefer to do leisure activities and browsing than shopping. In other words, the longer it takes a person to get to the center, the more likely he/she is to spend more time in the mall and do more leisure activities. It is also found that there is a significant relationship between the variety of shops in a retail center and the motivation of customers attending a retail center that encourages people to spend their leisure time in retail centers.

## About the authors

Sepideh is currently the PhD Candidate in Sustainability and Urban Regeneration program at the Universidad Politécnica de Madrid-UPM ,and she is doing her research on "The impact of shopping malls in High-Streets: the case of existing new urban development in Madrid". Saeed Nosratabadi is a Ph.D. candidate in management and business administration at Hungarian University of Agriculture and Life Sciences. His main research interests are business intelligence and food sustainability. Farshid Aram is currently the PhD Candidate in Sustainability and Urban Regeneration program. He has invented special software (Aram Mental Map Analyzer) to measure the citizens' mental cognition from urban spaces through mental maps (www.AramMMA.com). Amir Mosavi is a data Scientist, recipient of the Green-Talent Award, UNESCO Young Scientist Award, ERCIM Alain Bensoussan Fellowship Award, Endeavour-Australia Leadership Award, Humboldt Prize, Campus France Fellowship Award, Campus Hungary Fellowship Award, and Research Fellowship award of Institute of Advanced Studies.

**Public Interest Statement**

Retail Centers try to provide an environment in which customers can spend more time in the mall and at the same time enjoy spending time in the mall. In order to increase customer comfort and satisfaction, shopping malls also added recreational activities to the centers. The present study tried to investigate the factors that cause a customer to spend more time on these recreational activities in a shopping center. The findings of this study show that the distance that a customer spends to reach the shopping center and also the variety of shops in the shopping center are two main factors affecting customer behavior. In other words, the farther a customer comes to the mall, the more time he spends on the mall's leisure activities. In addition, it was found that the greater the variety of shops in a mall, the more time customers spend in the mall's leisure activities.

**1. Introduction**

According to the united nations' report in 2017, more than half of the world population predicted to live in urban areas by 2050 (UN DESA, 2017). Hence, making our cities more effective, equitable, and sustainable has been the focus of interdisciplinary research during the past half-century. Additionally, challenging urban areas to be more competitive with a focus on people's welfare in a sustainable development framework is a crucial issue (Coutard, 2008; Vasconcellos, 2014). Urban sustainability has been connected with keeping balanced and adhering retail systems set up in a great diversity of facilities, shopping spaces, and places (Cachinho, 2014; Moretti & Fischler, 2001). Since retailing has always been an urban activity par excellence, the services, and facilities that offer can be the main factors of their attraction. Nowadays, the relationship between city developments and retail has a little in common with the past. According to wide and deep changes, most cities have experienced the retail revolution, despite the differences that can be seen among the urban settings of different countries with different cultures. The effects of this revolution can be seen almost everywhere (Biao et al., 2010).

Along with urban development, the social dimension is a reflection of human development (D'Auria et al., 2018). This has been demonstrated as progress toward enabling all human beings to provide their essential needs, achieve an appropriate level of comfort, welfare, and share fairly in sustainable development opportunities. Sustainable development opportunities also composed of three topics: environmental, social, and economic evolution (Caulley & Bayley, 2012; D'Auria et al., 2018) (see Figure 1).

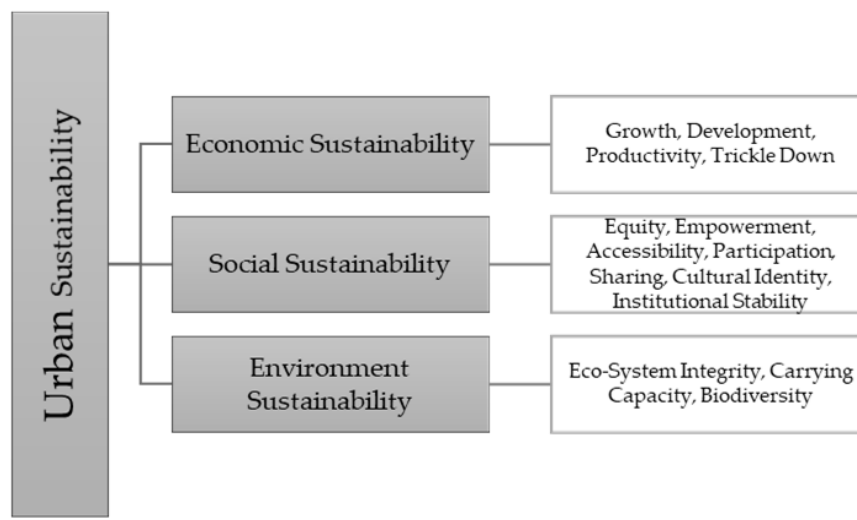

**Figure 1.** The paradigm of sustainable development in Agenda 21.

In 1987 the Brundtland report advises that to have powerful and dynamic communities, citizens' most foundational needs should be satisfied and enhanced (Visser & Brundtland, 2013). Shopping and retail activities are the necessity of economic and social activities that increase the vitality and livability of urban public life (Bİrol, 2005).

Designing retail centers can play an important role in creating the future of sustainability. Businesses can also simplify the systematic integration of environmental, economic, and social parameters in the framework of new and more sustainable models of production, marketing, and usage (Han et al., 2019). Regarding social sustainability, the built environment and leisure are viewed as required factors (Gozen Guner Aktas, 2012). Therefore, the importance of retail centers as retailing formats has become progressively remarkable. They have become central in customers' shopping pattern and have become more than merely a center to purchase (Sinha & Uniyal, 2005). Accordingly, for justifying deeper research on retail centers, the social factors can be one of the major criteria. For instance, leisure shopping is a browsing activity that is characterized by looking for pleasing experiences and stimulate the senses rather than a desire of purchasing (Bäckström, 2011).

There is a wide range of definitions for the word leisure. It can indubitably be in front of work or saying it as a synonym with entertainment which itself can be explained as effectively any satisfying experience (Torkildsen, 1983). Recently, leisure and recreational features of consumption have become more remarkable in the customer behavior research (Arnold & Reynolds, 2012; Hirschman, 1984; Jackson, 1991; Langrehr, 1991; Ritonga & Ganyang, 2020). Shopping is an activity that probably even be one of the preferred entertainments and a preferable activity of choice. Accordingly, it is a form of leisure activity (Bloch & Bruce, 1984; Codina et al., 2019). The main purpose of retail centers was once shopping, but now leisure activities are becoming integrated gradually (Kim et al., 2020). Customers in contemporary societies are spending more time on recreational activities, as such the originators of retail centers are trying to find a way to have leisure activities besides their commercial role (Aish & Marwaee, 2016; Torkildsen, 1983; Aish & Marwaee, 2016). Consumers of any ages prefer to spend more time in retail centers (Ewen, 2017; Martin & Turley, 2004; Taylor & Cosenza, 2002). Thus, retail centers have become community centers, and shopkeepers trying to offer more recreational attractions than before (Bloch et al., 1994; "Book Reviews," 1986; Guiry et al., 2006; Prus & Dawson, 1991; Tangherlini, 2011). Leisure activities can be the key to the success of retail centers since they can make an interesting and exciting experience for customers (Solomon, 2012). The diversity of the shops and their product assortment also encourage consumers to purchase more ("Book Reviews," 1986; Haynes & Talpade, 1996). Recreational activities and finding a place to offer recreational services are pivotal to retail center shopkeepers. Given that, they must create an interesting and entertaining atmosphere for customers that provides for their demands and extends their visiting time (Fox et al., 2004). Consequently, retail centers for social indicators through open spaces have a prominent value, which allows purchasers to have social contact (Lalović & Zivkovic, 2018; Sit et al., 2003). Hence, the assortment of retails, the retail centers' environment, and shopping involvement have a differential influence on excitement and desire to stay in retail centers (which in turn have an impact on patronage intentions and desire for shopping) (Ginsburg et al., 2007; Talarowski et al., 2019).

Numerous studies have shown that adding leisure activities to retail centers has become one of the obvious strategies of retail centers by which they can provide fun times to increase customer satisfaction. However, no research has been done to identify the driving forces behind customers' behavior in order to use the recreational activities of retail centers. In other words, although various research has been carried out on the effective factors that influence the purpose of going to retail centers, there are limited studies that directly identify customers' motivation for recreational activities. Therefore, this research particularly focuses on the factors that affect the recreational role of retail centers. Strictly speaking, the current study is conducted to bridge this research gap and it is intended to investigate driving factors behind the social role of retail centers on recreational activities. The article is arranged as follows: the key literature that discusses the social role of the retail center as a form of entertainment within the field of retailing briefly reviewed. This is followed by methodology and a showing of the results from the empirical study and finally, conclusions are derived, and future research is proposed.

## 2. Research Background

*2.1. The role of retail centers*

Shopping activity has been seen as a rapid global transformation during the past half-century. The development of different types of commercial-related sectors, in particular, following with an attitudinal change to consumption itself from the necessity to leisure (Wakefield & Baker, 1998). The retail centers which first became known in the United States have become the principal part of the contemporary lifestyle. It has been developing since the early 1920s by introducing changing patterns of shopping as well as social and recreational activities. Malls are also becoming places where shop keepers can touch their users in a fascinating ambiance (Mou et al., 2018). They have been defined as a group of retails and other commercial units that are evolved, organized, owned, and run as a single possession (Perera & Sutha, 2018). Consequently, nowadays we have been faced with changing the consumption pattern because of globalization, the international incorporation of markets for goods and services distinguishes the modern world economy. This process is known as globalization is providing more countries with opportunities to benefit from higher standards of living, but at the same time is making pressure on societies to adapt to their traditional practices (Devarajan, 2020).

Globalization makes the behavior of consumption progressively similar from one country to another. Critics of globalization often discuss that the strong pressure of market integration is forcing the world towards increasing homogeneity. These changes have also vastly affected retail centers and forced establishment and architects to build different types of retail centers for answering the various types of customers' needs (Kapustin, 2009). Retail centers can be comprehended from the evolutional perspective of consumption revealed as 'space of consumption' which all of these spaces have similar basal characteristics and also use similar methods to entice users (Mullins et al., 1999). Accordingly, to attract more customers and lengthen their stay in the retail centers, investors added various leisure venues and facilities to Retail centers (Jackson, 1991; Moretti & Fischler, 2001). However, it is not a new tendency and started many years ago but recently has accelerated (White, 2010). Considering studies in this field, we can see six periods of evolution in the conception of customers' recreational activities in shopping centers; 1930: the passive browsing and effect of capitalism on consumption pattern, 1950: the targeted Browsing and clarifying Spectrum of users, 1970-80: establishing retail with multi-purposes, 1990: adding leisure activities and popularized them, 2000: focusing on customers' behavior to make the retail centers more cognitive, and at present focusing on recreational activities, browsing, and targeted trip (Aram et al., 2020; Ilyin & Choi, 2017; Solomon, 2012) Shopping centers can affect recreational activities due to their size, number, and type of users they serve (Crawford, 1992; Eckert et al., 2013) (See Table 1).

**Table 1.** Europe shopping Centre Classification and Typical Characteristics.

Source: International Council Shopping Center, 2004 (ICSC)

Traditional shopping center

| Size | Concept | Typical GLA Range (Sq.M.) | Typical Type of Anchors |
|---|---|---|---|
| Very large | Classified by size, can be either indoor or outdoor, an all-purpose design | 80,000+ | Supermarkets, department stores, hypermarket, general goods store, Entertainment, Cinema |
| Large | Classified by size, can be either indoor or outdoor, an all-purpose design | 40,000-79,999 | Supermarkets, department stores, hypermarket, general goods store, Entertainment, Cinema |
| Medium | Classified by size, can be either indoor or outdoor, an all-purpose design | 20,000-39,999 | Supermarkets, department stores, hypermarket, general goods store, Entertainment, Cinema |
| Small | Comparative, Centers include shops typically purchasing clothes, home furnishings, electronics, general goods, accessories, and other discretionary merchandise | 5,000-19,999 | Not usually anchored. |
| | Convenience-Based, Centers include shops that purchase crucial goods | 5,000-19,999 | Typically linked with a grocery store Additional stores usually include pharmacy, convenience stores, and shops that meet other needs |
| **Specialty Shopping Centre** | | | |
| Retail Park | Also known as a power center. A consistently designed, planned, and managed open and large-scale specialist retailers that are mostly freestanding | Large: 20,000+<br><br>Medium: 10,000-19,999<br><br>Small: 5,000-9,999 | Usually linked by discount department stores, warehouse clubs, off-price shops, or another category |
| Factory Outlet | Outdoor and/or enclosed center that comprises manufacturers' and retailers' outlet stores selling brand | 5,000+ | Generally, not anchored, although some brand-name stores may act as magnet shops |
| Theme-Oriented Centre | Leisure-Based - A consistently designed that includes some retail units and typically focuses on a narrow but deep selection of goods within a particular retail class | 5,000+ | Usually linked with a multiplex cinema |
| | Non-recreation-Based - A consistently planned that includes some retail units and typically focuses on a narrow but deep selection of goods within a particular retail class. | 5,000+ | NA |

## 2.2. Leisure time and opportunity in retail centers

Leisure has vast definitions, contrasting leisure with work or mentioning it as synonymous with recreation (which itself can virtually be seen as any satisfying experience) (Bäckström, 2011). Leisure times can be a time of relaxation, comfort, calmness, and free time that anyone uses differently. The following table briefly refers to the provision of leisure from different viewpoints (See Table 2).

**Table 2.** Leisure's Definition.

| | |
|---|---|
| Cambridge Dictionary | - The time when you are not working or doing other duties |
| Collins Dictionary | - Leisure is the time when you are not working and you can relax and do things that you enjoy. |
| Oxford dictionary | - The Time of someone who is not working or occupied, |
| George Butler | -Healthy recreational activities that people do in their free time cause mental expansion, growth, and character improvement. |
| Stanley Parker | - Leisure is the remaining time that an individual has after all other activities |
| Max Kaplan | -A pleasant memory, a kind of freedom from duties |

In the 1950s when the malls were born, we rarely could see the entertainment and recreational activities in retail centers or malls. Between 1895 and 1920 the recreational activities such as movie theater, amusement parks, were built in the downtown of the US. Nowadays retail centers are not only being used merely for shopping, but their new role is also motivating users towards entertainment and spending leisure time (de Los Santos et al., 2012). There are also four main relationships between shopping and leisure, shopping as purchasing, shopping for leisure, shopping and leisure, and shopping as leisure. Both with the definitions of leisure intrinsic in the approaches, as well as with various properties of shopping facilities, such as their size and complexity that are loosely related; see Figure 2, which is an adaptation from (Howard, 2007).

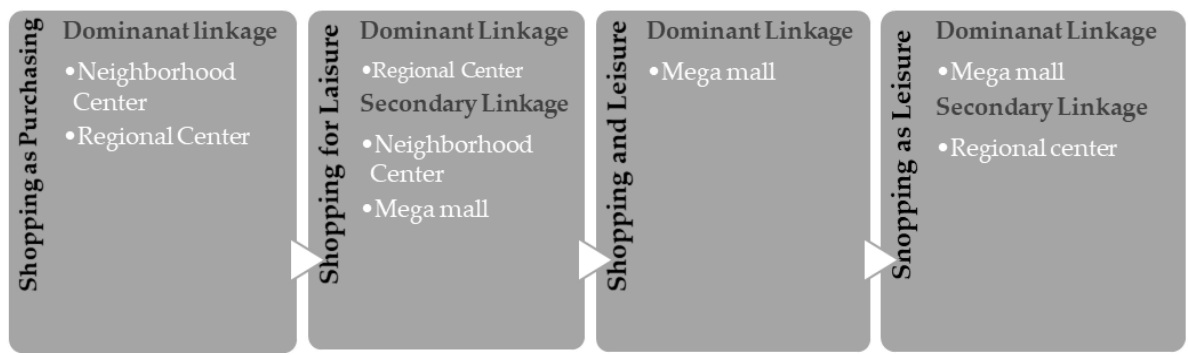

**Figure 2.** Linkages among the nature of leisure, relationships between leisure and shopping, and the scale of retail facilities.

## 2.3. Retail centers and their social role

For the first time in the United States, chain stores increased in the late 1920s, this type of market flourished and turned into supermarkets with discounts in the 1930s (Ewen, 2017; Gory & Jacobs, 1985).

In 1956, with the founding of the first retail center in Southland by Victor Groningen, his utopian assets, this retail center became a gathering place for civic and social meeting place (Bailey, 2015). This trend continued in the following years, and many retail centers became indoors in the 1970s such that the attitude of retail centers sought to add side-effects such as food stores, cinema, and theater ("A Brief History of the Market," 2015). Almost everywhere in the United States, a large number of teenagers set their appointments in the suburban vicinity of the retail center. In research in 1985, 63 % of the retail centers' users came there once or twice a week, and most of them spent one to five hours in these spaces. They rarely just went shopping alone and usually went out in these spaces with their friends. Only half of the people said that they do shopping as their goal of coming to retail centers (Zhang et al., 2019).

Nowadays we can leave our children to play in malls' playgrounds while we are doing shopping or other activities since most retail centers offer more recreational activities. Hence, retail centers that initially held a more commercial role are now considered places of recreation (Frączkiewicz, 2013).

The sphere of retail centers affects shopping habits. There are various factors which could define the utility of retail such as, travel distance (Li et al., 2021; Severin et al., 2001) travel time (Brunner & Mason, 1968) accessibility (Blut et al., 2018) scale (Ushchev et al., 2015) variety of shops (Dijkstra et al., 2011), and the number of stores (Gorter et al., 2003). Retail centers can also explore place attachment (Kusumowidagdo et al., 2015). There is a relationship between social and physical indicators which form a place with visitors' behavior (Najafi & Shariff, 2011), while indicators can generate the place's characteristic (Foote & Azaryahu, 2009). Spacious shopping places have increasingly become a place where we do not only do shopping but also, we can spend our free time. The shopping atmosphere, with the colorful alley, which allows us to freely walk around them, even in bad weather conditions. They also provide more activities like shopping, dining, entertainment, cultural events which are gradually moving from city centers to large places. These days, retail centers have been used in different ways by different groups of people (see Table 3) (Frączkiewicz, 2013; Storey, 2014).

**Table 3.** The social role of retail centers.

| | |
|---|---|
| Trade Function | The most noticeable one. The galleries are mainly shops offering Famous products, made by popular brands. |
| Catering | In retail centers, the place can mostly find Fast food restaurants which reveal the nature of a typical consumer those places |
| Social | Large retails have increasingly become a place where we can not only do shopping or eat but also spend our free time. |
| Recreational | In most modern retail centers, there are large chain gyms open, Beauty services, where in addition to the previously mentioned social roles of malls, you can also walk with shopping and dining opportunities, work on your appearance. The many galleries multiplex cinemas, entertainment centers, clubs, and even pubs are located in most modern retail centers. |

In this regard, this study examines social events and how these spaces are transformed into public spaces in one retail center in Tehran and one retail center in Madrid. In Iran traditionally markets have always been the venue for social events (Sajadzadeh & Haghi, 2018). With the arrival of modernism in Iran, major changes were made to the pattern of designing shopping spaces. The establishment of Ferdowsi's stores in 1957 and Cyrus's shops in 1970 and the expansion of them in the city created a new style of shopping in Tehran (which was influenced by emerging atmosphere of Tehran's lifestyle) (Afkan & Khorrami Rouz, 2015). With the outbreak of the Islamic Revolution and the Iran-Iraq war, the process of extending retail centers was stopped, and after that, they were sold or renamed. In the 1990s, new chain stores such as Shahrvand and Refah were established in Tehran (Afkan & Khorrami Rouz, 2015). In the design of retail centers and chain stores of the 90s, there was no place for life. These types of markets were mostly passageways. the leisure activities were not the main issue of these spaces. In the early 21st century, major urban projects were conducted in Tehran. In that period, the municipality encouraged the contractors to build high rise buildings, towers, and retail centers. Many retail centers were built in Tehran during this decade such as Golestan Shahrak-e Gharb, Ghaem Tajrish, and these spaces turned into a youth meeting point. Starting construction in the last decade, from 2010, and adding more services such as food court, cinema, playground, a new type of commercial-recreation spaces pattern in Tehran formed (Habibi & Mahmoudi Pati, 2017).

In Spain centers and supermarkets have experienced an important development since their appearance in Spain in the 70s. In the 80s the first retail centers of Spain were inaugurated with lots of facilities, shops, offers, parking spaces. A case in point, the La Vaguada, and Parque Sur where the retail centers were established in the 80s in Madrid. In the 90s the trend of retail centers was devoting more space to leisure. Finally, since 2000, we have seen the retail shops saturated with cinemas and comprised of sports. The most representative example is Xanadu, which opened in 2003 in Madrid (López García De Leániz & Míguez Iglesias, 2017). In summary, Spain has three periods of retail shop transformation; up until the 80s hypermarkets periods; from the 80s until the 90s (the advent of retail centers); and from the year 2000 (diversity of retail centers has been heightened). Recently, retail centers in Madrid have not only been considered a retail center but have also become a leisure and entertainment center. These spaces are not just meeting point for youth but can be used as a recreational destination, in which they can spend their full day buying, going to the cinema, having fun, and eating (Marín de la Cruz,2013).

Many articles have been studied on the role of retail centers on economic and environmental sustainability. However, rarely some researches have been worked on the effect of retail centers on social sustainability and the impact of retail centers 'social role on consumers' behavior and their shopping pattern. Accordingly, the lack of study on this new role of the retail center and its impact on the customers' recreational activities in terms of social sustainability has seen as a gap that needs to be worked on it. On the other hand, similar studies that have been conducted in this field mostly studied the retail centers located in the same city or country. However, in this study, it has been attempted to examine the retail centers that are located in different countries with a dissimilar consumption culture. Consequently, this study aims to investigate the effect of retail centers 'new role as places for interactional and recreational activities on customers' presence reason in these centers and its relation to social sustainability. This study focuses on the perception of retail centers from the customers' point of view via studying dependent and independent indicators to determine the role of these centers as a place that provides recreational activities along with shopping. Hence it was hypothesized that:

**Hypothesis 1**: The longer it takes a person to get to a retail center, the more likely he/she is to spend more time on the recreational activities.

**Hypothesis 2**: The greater the variety of the shops in a retail center, the more people are willing to spend more time in that center and do recreational activities.

## 3. Materials and Methods

*3.1. Data collection*

The current research in terms of the purpose is an applied comparative study, and according to Collier (Ramrattan & Szenberg, 2019) a comparative method provides a foundational tool for analysis and producing an insight. A mixed-methods approach has been used for the study, to fully explore the relationship between customers and retail centers and to ensure that the most important factors were identified and properly conceptualized for the data collection. For the qualitative factors, a series of semi-structured interviews were run to identify the components related to the relationship between customers and the retail center's social role (Horton et al., 2004). A questionnaire was then created and pre-tested on the interview subjects. Two case studies were conducted in two analogous retail centers, one in Tehran, Iran and the other in Madrid, Spain

*3.2 independent and dependent variables*

In this study, we used two independent variables, first, the travel time, indicating the time takes customers for attending the retail centers, and, the motivation of customers for their attendance in these retail center. The variety of shops in the retail centers is mentioned as a dependent variable, since it can be almost one of the main factors for attracting customers, the variety of shops acts as an influential factor on dependents factors of this study (Wakefield & Baker, 1998).

*3.3 Data analysis*

The structural equation modeling (SEM) is applied for data analysis and for hypotheses testing, in this study. In the SEM, one of the most important steps is to express the model (Nosratabadi et al., 2020). Model expression in the SEM is about clarifying relationships among variables. After model expression, free parameters should be estimated. There are many iterative methods, such as partial least squares (PLS), generalized least squares (GLS), and maximum likelihood estimation (MLE), that are able to estimate the model fit (Ulman & Bentler, 2003). The current study applied PLS method using SmartPLS 3.2.8 for testing the model. In the SEM, models constitute of a structural model and a measurement model. Indeed, the relationship among main variables, which are also called latent variables, is called the structural model and the relationship between the observable variables and corresponding latent variables is called the measurement model (Sarstedt & Cheah, 2019). In other words, latent variables cannot be measured directly, therefore, the observable variables are used to measure a latent variable. Equation 1 is a regression model can measure observable variables so as to estimate the corresponding latent variable:

$$x_{pq} = \lambda_{p0} + \lambda_{pq}\xi_q + \epsilon_{pq} \qquad (1)$$

In Equation 1, $\xi$ represents the latent variable, $\lambda$ is the loading factor, and $\epsilon$ shows the error in the measurement process.

To test the degree of correlations in a measurement model, the average variance extracted (AVE) is used. AVE, in fact, evaluates the convergent validity. A high convergent validity means that the latent variable gains most of the variances from its own observable variables, not from the error measurement. AVE can get a value between 0 to 1 that a value greater than 0.7 is more desirable and values higher than 0.5 consider acceptable. Equation 2 shows how to calculate AVE:

$$AVE = \frac{\sum \lambda_i^2}{\sum \lambda_i^2 + \sum \Theta_{ij}} \qquad (2)$$

Where $\lambda i2$ is the factor loading and $\Theta ij$ is error variance. The error variance is estimated in Equation 3:

$$\Theta ij = \sum 1 - \lambda_i^2 \qquad (3)$$

In the SEM, along with Cronbach's alpha, the composite reliability ($\rho c$) is used to test internal consistency. These two metrics represent the reliability of a model. Equation 4 is used to calculate the composite reliability:

$$\rho c = \frac{(\sum \lambda_i)^2}{(\sum \lambda_i)^2 + \sum \Theta_{ij}} \qquad (4)$$

A dependent variable is called endogenous latent variable and an independent variable is named exogenous latent variable, in the SEM. Equation 5 is for calculation of an endogenous latent variable:

$$\xi = \beta_{oj} + \sum \beta_{qj}\xi_q + \zeta_j \qquad (5)$$

In Equation 5, $\xi$ is an endogenous latent variable, $\beta_{qj}$ is the path coefficient between the q exogenous latent variable and the j endogenous variable. $\zeta_j$ represents to the error in the inner relation.

*3.4. Study Site*

Kourosh complex, which is located in Tehran, Iran, and the Las Rosas retail center, which is located in Madrid, Spain, are two case studies of the current study. These two retail centers are almost similar in terms of characteristic and functional, since both are community malls (Devarajan, 2020; Kapustin, 2009) and have been built in the current context, and also were established in middle- class districts that are far from the city center were selected as evaluation models (see Table 4).

**Table 4.** Similarity and disparity of las Rosas retail Center& Kourosh Retail Center.

| Similarity | | |
|---|---|---|
| **Kourosh Retail Center** | **Las Rosas Retail Center** | Retail Centers |
| Character | Size | Las Rosas Retail center |
| Location | Function | Kourosh Retail Center |
| Disparity | | |
| The similarity of two Retail centers Disparity of two Retail Centers | | |

3.4.1 Tehran: Kourosh Complex

Tehran, with a population of 8.70 million, is the largest city in Iran. It contains many retail shops, one of which is Kourosh Complex, built-in 2009. With an area of approximately 9500 square meters and a population of 858.346, it is located at the intersection of Sattary highway and Payambar street in district number 5 in the northwest of Tehran, near the tourist area. The total number of retail shops is 550 units, this retail center was constructed in a middle-class neighborhood, which was constructed around the same time.

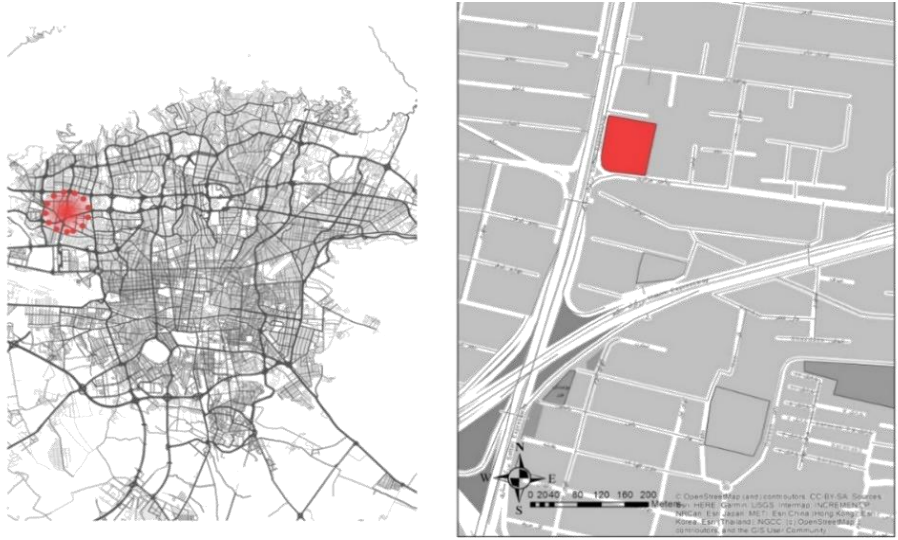

**Figure 3.** Location of Kourosh Retail center in Tehran.

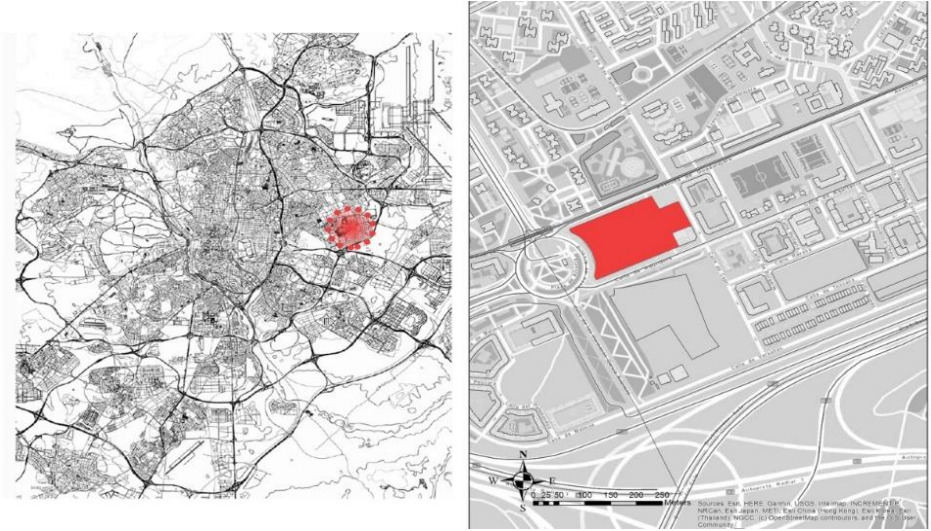

**Figure 4.** Location of Las Rosas Retail center in Madrid.

Kourosh Retail center is a complex of commercial, cultural, and entertainment, including a cinema, campus, kindergarten, cafe, VIP restaurant, and a supermarket. According to previous findings, the major item sold currently is clothing. In addition to high priced brands, Kourosh complex offers apparel for middle incomes as well. Apart from the shops, shoppers can enjoy spending time in the food courts, restaurants, coffee shops, game lands, and cinema complexes available on the top floors of the mall. Additionally, the parking lot is quite spacious, with several floors underground. Measurements in Tehran were conducted on the 12th of April 2019 during peak hours of Friday that is the weekend of Iranian, from 20:00 to 22:00. According to the Cochran formula, in total 150 questionnaires were collected from Kourosh retail center.

3.4.2 Madrid: Las Rosas Commercial center

Madrid, Spain's central capital has a population of 3.17 million and more than 100 retail centers. One of these centers, Las Rosas, is located in the San Blas-Canillejas district on Avenida de Guadalajara with a population of 158.9 thousand, It was built in June 1998 with an area of 110.000 sqm2, It has two access entrances, the main entrance at Avenida de Guadalajara, and the second one by the street of Aquitaine (Madrid city council,2016). The center consists of two commercial levels with 61 stores, a cinema, restaurants, and markets. The first floor is primarily occupied by Carrefour and Mercado Clasico, while the second floor is made up of restaurants and the cinema. Las Rosas was established in one of the oldest districts in Madrid and is made up of low-or middle-income residents. The retail center itself was built along with the neighborhood and most of its customers are residents. By using The Cochran formula, 150 questionnaires were obtained from Las Rosas retail center on the 2nd of March 2019. The responders have answered the question during the peak time of Las Rosas shopping center from 18:00 to 20:00.

**4. Results**

The outcomes of this study derived from the evaluation and analysis of 150 questionnaire data distributed across Kourosh retail center and 150 questionnaires were collected from Las Rosas, using SmartPLS 3.2.8. In this research, to investigate the social role of retail centers as places for interactional activities on urban social sustainability, three variables were considered, two independents and one dependent, the travel time to retail centers and the variety of retail centers' shops as independent variables and the motivation of customers to attend retail centers as a dependent variable. Information regarding the travel time variable was collected by questionnaire to examine the time that shoppers spent traveling to the retail centers studied. This variable provided information about the retail center's attraction. This variable was considered as a factor that demonstrated shoppers' attitudes towards attending the retail center. However, one exception is shoppers with a short travel duration, which generally applies to residents who live in the vicinity of the retail centers. The second variable, customers' motivation to attend retail centers, was divided into three main categories: shopping, leisure, and browsing. The questions in this section were asked by a researcher inside the retail centers to clarify the reason behind shoppers' presence. By analyzing this indicator, the main stimulus of customer presence was revealed. The variety of shops as a dependent variable collected by field study has been categorized into five categories, restaurants and leisure activities ("A Brief History of the Market," 2015) fashion and accessories, technology, supermarkets, and beauty services that will show us the main tendency of each retail center attracting their customers.

*4.1. Measurement Model*

4.1.1 Tehran: Kourosh commercial center

The extent of the attraction towards the Kourosh retail center was analyzed according to the customers' travel time, which was obtained from the questionnaire and analyzed by Eta Squared.

Duration of fewer than 10 minutes was associated with residents who lived in the retail center's vicinity and were not considered useful for the attraction variable. However, a travel time of more than 10 minutes was considered an effective factor in the retail center's attractiveness. The results of this research indicated that individuals with a travel time of between 0 and 10 minutes and between 11 minutes and 1 hour have the highest frequency of attendance at Kourosh complex, with 26% and 74%, respectively. Additionally, it is of interest that the highest percentage of shoppers present at the Kourosh center was located outside of the vicinity. This center was of special interest in this study due to its commercial-recreational features. According to Figure (5,6), approximately 70% of people visited this venue for more than 20 minutes, with 37% desire of doing recreational activities and only 18% of them desire to go shopping. The statistical results shown in Table 7 indicate significant effects (i.e., p values are below .05) for the motivation and the travel time. Note also that the travel time duration x motivation interaction accounts for 13.9% (i.e., $\eta^2$ value is more than 0.09) which shows a strong interaction effect between the travel time and the motivation. The Kourosh Retail Center is a local retail center that was established to meet the needs of the 5th district of Tehran. Nevertheless, these data demonstrate that in addition to meeting the essential needs of the residents, this center is considered a hub of leisure activities as well. Figure (7) indicates that 46.7% of the respondents come to this center for leisure activities and only 36.7% came for shopping. Meanwhile, the percentage of individuals who browsed the stores and did not spend money (16.7%) is of interest. Due to Figure (8) that demonstrates the high variety of shops in this retail center the correlation of these two aspects, the variety of shops and desire of spending recreational activities, Table 7 shows that they have a significant correlation together.

4.1.2 Madrid: Las Rosas Retail center

The study was conducted four different times using a questionnaire that was completed by 150 respondents. The time corresponding to 0 - 10 minutes applies to residents who use the retail center to supply daily necessities, and therefore was not considered an effective factor. However, the period from 10 - 60 minutes is considered a significant factor. Although this retail center is a neighborhood retail center, the results showed that individuals who spent 0 - 10 minutes to use this space comprised 41.33% of the total, while those who spent 10 - 60 minutes comprised 58.66% (see Figure 5). On the other hand, the visitors who attended the retail center's aims were divided into three categories: shopping, leisure, and browsing. According to the data obtained in this study, recreation activities in a period of 11:00-1:00 Hrs. were the most frequent (42.0%), followed by Browsing (10.0%), and Shopping (7.0%) Figure (6).

As regards the obtained data, the statistical results have been shown that (see Table 7) there is a significant correlation between the motivation and the travel time. Note also that the travel time and the motivation interaction account for 22.1% that shows the interaction effect between these two aspects. Concerning the second hypothesis, the acquired data Figure (7-8) analyzed by Pearson correlation coefficient (see Table 7), the results show that there is a significant correlation between the variety of shops and customers' motivation, which means the more variety of shops increasing the desire for doing more recreational activities.

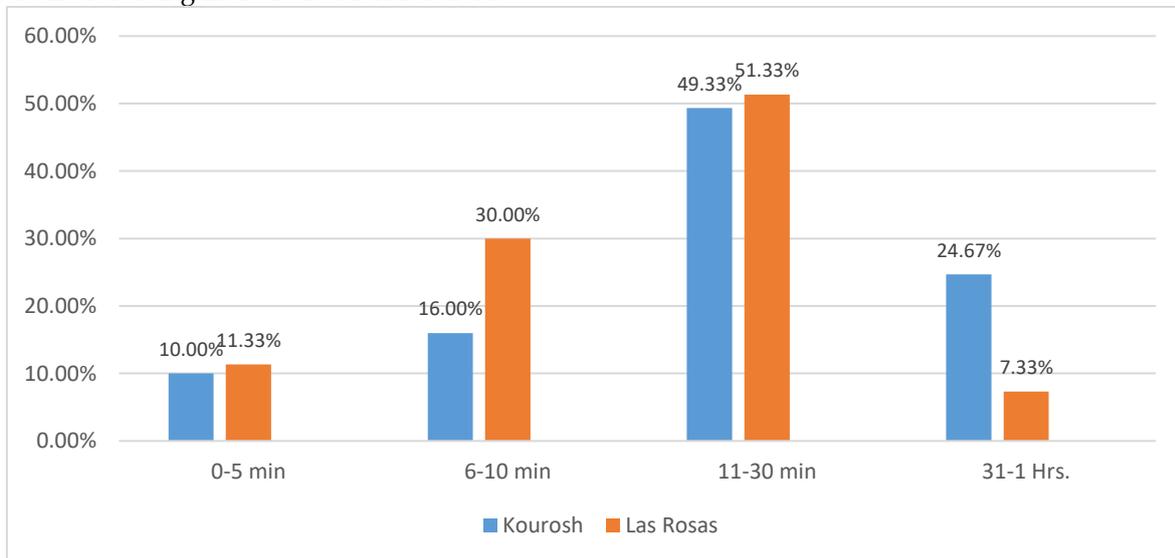

**Figure 5.** Kourosh and Las Rosas retail center's travel time duration.

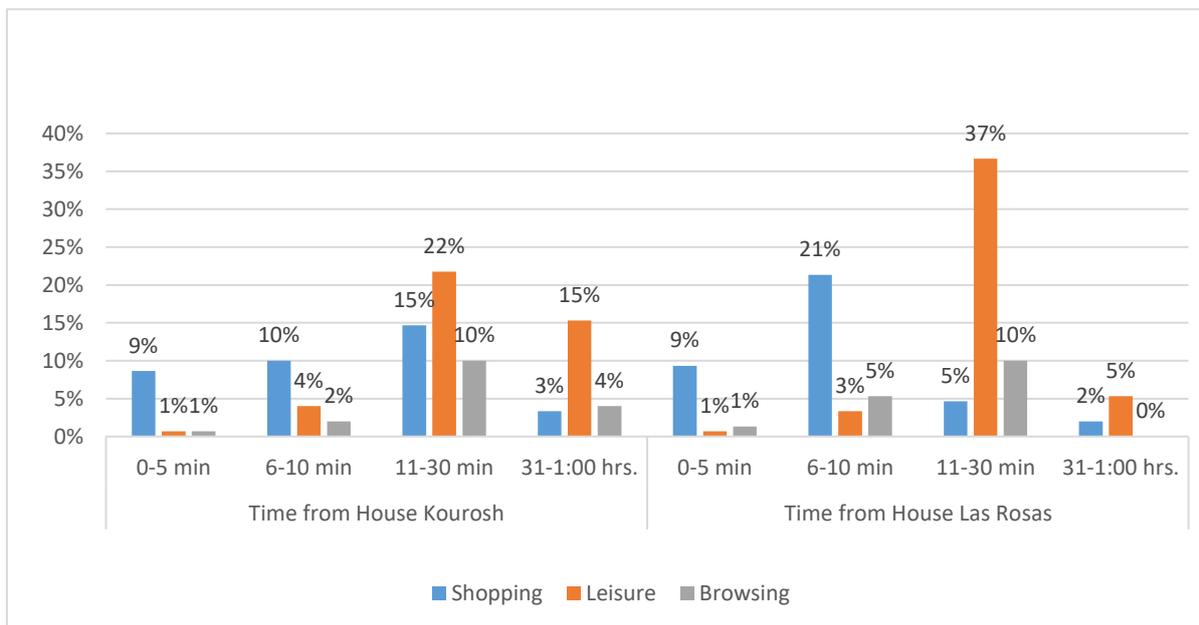

**Figure 6.** Kourosh & Las Rosas retail centers 'travel time and shoppers 'desire.

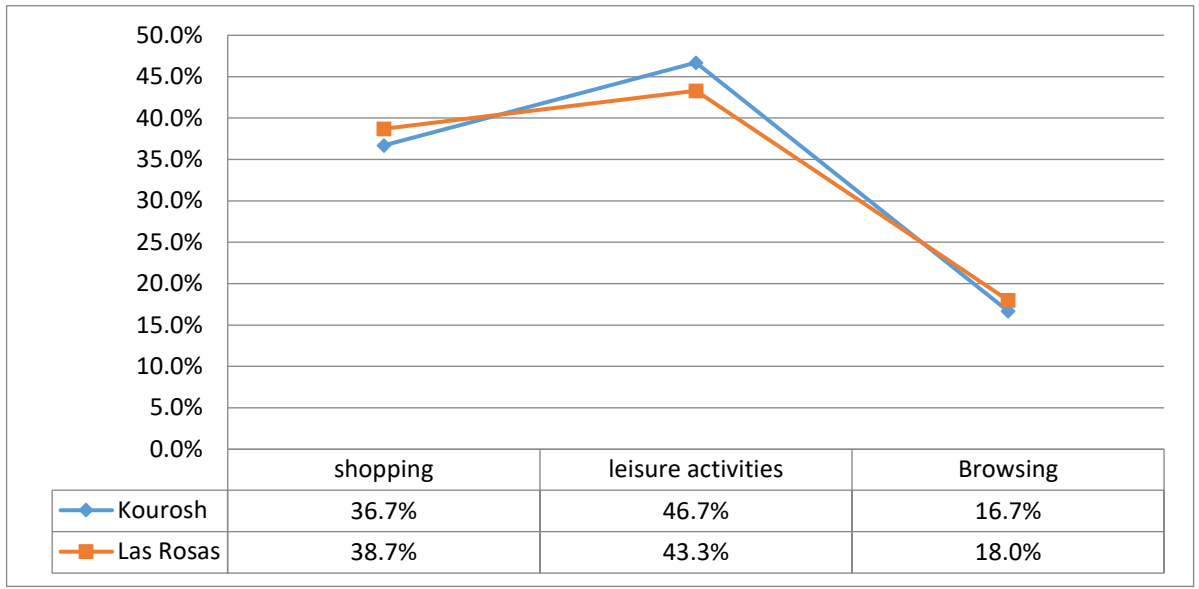

**Figure 7.** Kourosh and Las Rosas Retail Center's Motivation.

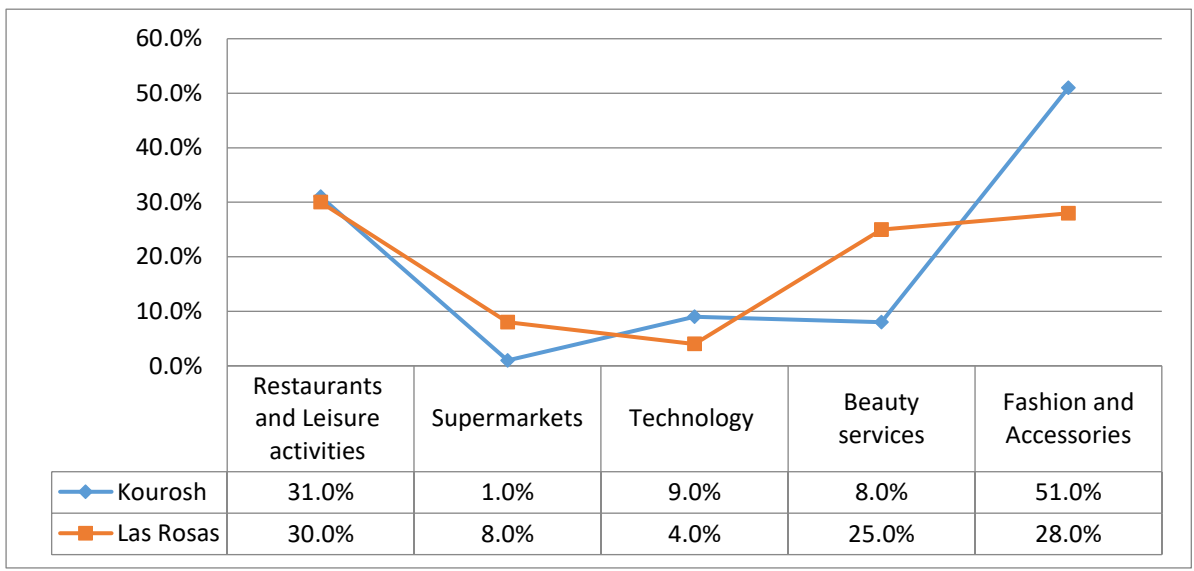

**Figure 8.** Kourosh and Las Rosas Retail Center's Shops' Variety.

*4.2 Testing Hypotheses*

The AVE was used to assess the convergent validity of the data collection tool of this study, namely the questionnaire, and Table 5 shows that the value of AVE is higher than the acceptable limit for all variables, i.e., 0.5. The AVE is equal to 0.688 for time interval and higher than the desired value, i.e., 0.7, for the variety of shops and the motivation. This means that in all measurement models of the proposed model of this study, more than 50% of the variances of all three variables are defined by the observable variables of the corresponding variable, which indicates the high convergent validity of the questionnaire. Cronbach's alpha and composite reliability (CR) were used to measure the reliability of the measurement model. Table 5 represents that all Cronbach's alpha values and composite reliability for all three variables are above the desired value of 0.7, which indicates the high reliability of the measurement model and data collection tools in this study.

Table 5: Validity and Reliability of the variables

| Variables | AVE | Cronbach's alpha | CR |
|---|---|---|---|
| Travel Time | 0.688 | 0.74 | 0.941 |
| Variety of Shops | 0.832 | 0.91 | 0.918 |
| Motivation | 0.721 | 0.88 | 0.939 |

SmartPLS 3.2.8 was used to test the hypotheses and evaluate the structural model of the present study. In the SEM, three metrics of the coefficient of determination ($R^2$), loading factors, and path coefficients are considered to measure the model fit. It should be noted that the path coefficients and loading factors are acceptable if the corresponding T-values are significant at a level above 95% ($p \leq 0.05$).

Figure 9 is the output of SmartPLS 3.2.8 and illustrating the result of testing the proposed model of the study. The $R^2$ indicates what percentage of the dependent variable changes are determined by the independent variables (White, 2010). According to Figure 9, $R^2=0.787$ that means 78.8% of the variable of motivation is explained by the variables of time interval and shops' variety, which is a considerable number.

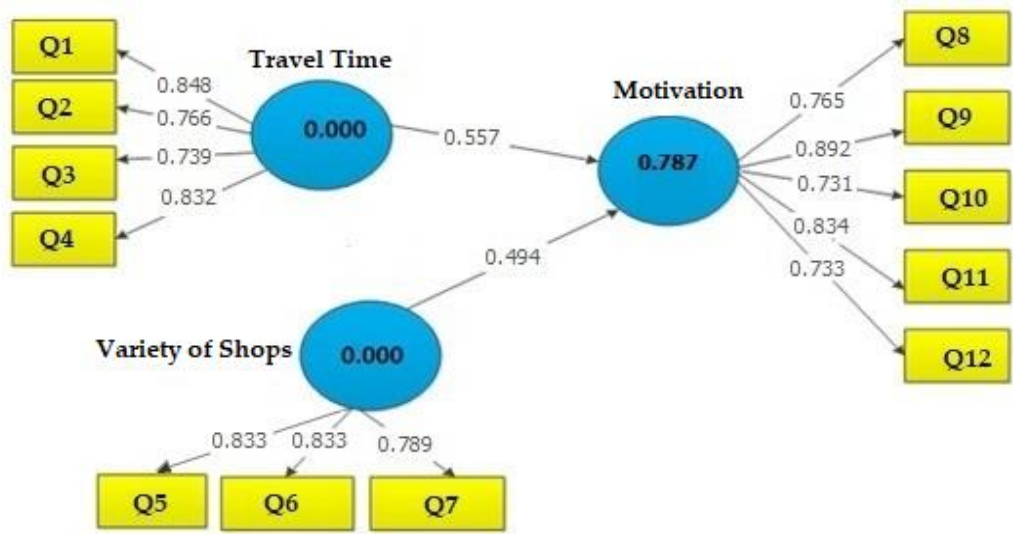

**Figure 9.** The results of testing the hypotheses: R-squares and Path Coefficients

The results of the significance test of loading factors are given in Table 6. Table 6 shows that all loading factors of all three measurement models are significant above 99% confidence.

Table 6. The T-values of the model test

| Variables | Travel Time | Variety of Shops | Motivation |
|---|---|---|---|
| Q1 | 6.404 | | |
| Q2 | 13.133 | | |
| Q3 | 10.171 | | |
| Q4 | 9.092 | | |
| Q5 | | 14.824 | |
| Q6 | | 18.571 | |
| Q7 | | 3.501 | |
| Q8 | | | 5.102 |
| Q9 | | | 5.8 |
| Q10 | | | 13.409 |
| Q11 | | | 11.837 |

|   |   |
|---|---|
| Q12 | 4.318 |

** all the loading factors are significant at the 99% confidence level

After testing and confirming the fit of the measurement models, the structural model was tested. The structural model refers to the relationship between the main variables of this research and in other words, tests the main hypotheses of this research. The result of testing the hypotheses and the output of SmartPLS 3.2.8 is presented in Table 7 and Figure 9. Table 7 shows that the path coefficients of the time interval to the variable of motivation is equal to 0.557 and this path coefficient is significant at the level of 99% (P≤0.01) that indicates that the corresponding hypothesis is confirmed. The path coefficients of the shops' variety to the variable of motivation is equal to 0.494 and this path coefficient is significant at the level of 99% (P≤0.01) that implies the second hypothesis of the study is confirmed as well.

**Table 7.** The results of hypotheses testing

| Hypotheses | Path Coefficient (β) | T-value | $R^2$ | Result |
|---|---|---|---|---|
| Travel Time → Motivation | 0.557** | 5.923 | 0.787 | Confirmed |
| Variety of Shops → Motivation | 0.494** | 8.065 | 0.787 | Confirmed |

** the correspond path coefficient is significant at 99% confidence level

## 5. Discussion

Regarding the results, shows that the Kourosh retail center is located in Tehran, and Las Rosas retail center is located in Madrid, along with their commercial role their social role is contemplative. Due to the study, three factors were examined, the travel time, the costumers' motivation to attend in these centers, and the variety of shops. The obtained research data demonstrate that most consumer's tendency to visit retail centers for leisure and entertainment activities is the main purpose and followed by engaging shopping. Results show that Kourosh Retail Center in Iran and Las Rosas retail centers in Spain have a similar function on clients' willingness to do entertainment activities such as doing sport, going to cinema or restaurants, by 46.7% and 43.3% respectively. Also, the results indicate that the desire of residents for browsing without a prerequisite purchase intention for each retail centers is 16.7% and 18.0% respectively, in total 63.4% of costumers' motivation in Kourosh retail center and 61.3% in Las Rosas have devoted to activities other than just purchasing products. (see Figure 7), In the meantime, two factors of the travel time and the variety of shops can be the main influential factors affecting customers' increasing propensity to spend their leisure time in these centers.

Various studies have been conducted on influencing factors that impact using retail centers, but the studies that directly mentioned the effects of the variety of shops on customers' motivation for recreational activities are limited. The researches that have been studied by Fennell (1978) Moschis (1976), suggested that the more complex or differentiated products in outlet retail centers lend customers more to user exploration and information seeking. The other research examined three retail centers in the United States in 1986 ("Book Reviews," 1986). In this study examined the relationship between age and income with the shop's variety, and finally, determined which types of shops are most attractive for which age groups and income levels. Simon Bell (1999) studied psychological factors that can affect customers' Image and attraction, via his study the shops' variety besides other aspects such as price fairness, visual amenity, convenience, and customer services strongly were associated with customers' selection of retail centers. Calvo Porral and Lévy Mangín (2018) studied pull factors for attracting more customers in Spain, they mentioned that the variety of shops is one of them and can affect attracting customers.

Despite the studies that have been done, they just revealed that the variety of shops is an effective factor in increasing the customers 'desire to purchase or go to retail centers and didn't deal with the particular effect of the variety of shops on increasing customers 'desire for recreational activities. Furthermore, regarding the first hypothesis, the results show that almost 51%- the travel time of

Kourosh' purchasers to get this center is more than 10 min (see Figure 7), compared to Las Rosas retail center this percentage is 52% (see Figure 6). In fact, regarding the variety of shops in both retail enters show that the main emphasis in both of these retail centers is on entertainment activities and because of variable involvement in the merchandise presented for sale it makes customers find these places pleasurable for browsing or window shopping (see Figure 8), which indicated that both of these factors are somehow related to costumers' motivation, therefore when the travel time is long and also the level of variety of shops is high, equally, the percentage of recreational activities will increase. In this case, few studies examined the travel time role as an impressive factor in retail centers, Louail et al. (2017) suggested that preserving main fundamental properties as the travel time at the same time can be increasing equity of spatial in cities. The study which has been investigated by Beiró et al. (2018) in Chile showed that the customers prefer to select the retail centers which take them less time and can get there easily. The results of Handy et al.'s research in the US (Handy & Clifton, 2001) revealed that the shopping center-goers prefer to go to centers which take them more time because they prefer to drive than a walk in this case the local shopping don't have any specific role for reducing the automobile, in this research the factor of time only was studied to show the effect of retail centers located on the type of transportation. The research which studied by Brunner and Mason (1968) in Ohio demonstrated that the driving travel time besides population density, purchasing power, and other factors can be influential can be one of the most important aspects to reach a retail center is extremely influential in deciding user retail center preferences.

In general, based on research done in this regard, it has shown that nowadays, shopping in retail Centers is widely considered a recreational activity and can be seen as a main factor competitive instrument in the retail industry. Tripathi, Roy and Mishra (2020) recently argued that going to the retail centers carries out more purposes than pure consumption and many people prefer to go to the retail centers rather than go anywhere else. Nisco and Napolitano (2006) found a positive connection between Entertainment orientation and performance outcomes of a retail center empirically. The research in the city of Guangzhou showed that the consumption activities of most of the consumers are mainly for recreational activities, besides, their space consumption and cultural experience in retail centers have more nonmaterial factors (Lin et al., 2010). The other research which is studied in India shows that nowadays in India for buyers, shopping is not like a chore but it has become a leisure activity, due to this changing the retail centers try offering more leisure activities to pursue more customers (Kuruvilla & Ganguli, 2008).

Despite all research done, they don't specifically mention the features that can improve this role of retail centers as a place for spending leisure time which can work as a successful one due to attracting more customers. Therefore, this study tried to survey the features which can have an effect on the number of visitors and also persuade them to stay more in retail centers, hence due to results of this study which provide for the research hypothesis, more precisely, the study gives a demonstration of the importance of retail centers' role as a place where provide Recreational activities apart from shopping activities and also disclose that retail center attractions mostly drive retail center arousal. In other words, the power of retail center to attract users may be explained by several indicators, the main factors which affect shopping arousal among shoppers concern recreational facilities (Rajagopal, 2009) all retail centers are to some level leisure centers which may be the destination of leisure trips, this issue was inspected via time duration and shopping motivation as an independent factor and shops' variety as a dependent one.

In the second hypothesis, it is was seen that the variety of shops is also one of the important issues to attract users, as mentioned in the latter part in both retail centers the highest percentage belongs to retails which offer the fashion, accessories and recreational services which attract shoppers for browsing and spending their leisure time in these spaces. Eventually consequence of these two hypothesizes are evidence of the retail variety may be the main factor for reveal the motivation of users and also the trip time duration can emphasize this motivation for spending more time on recreational activities which means that the more travel time that individuals spend, increase the percentage of leisure activities. Concerning the first hypothesis in association with most customers' motivation which is spending leisure time in these retail centers, the time it takes the 46.7% of Kourosh shopping centers-goers and 43.3% of Las Rosas Retail Center-goers to get the retail center demonstrated that the majority of individuals are from further distances and takes them more time to get retail centers, that shows that the longer trip can be one of the factors for willing to spend more leisure time and recreational activities. For future research, the new hypotheses involving eco-innovation, sustainable retail, and novel branding technologies (e.g., Palazzo, et al. 2021; Yousef, et al. 2021; Fakhredin, Foroudi, and Ghahroudi 2021; Sumrin, et al. 2021), would be beneficial for a comprehensive research and outcome

## 6. Conclusions

The current study investigated the research on investigating the social role of shopping centers on customer behavior regarding social sustainability. Findings of this study disclosed the interaction between the travel time and the variety of shops as independent variables and motivation as a dependent variable. In other words, it is revealed that the longer it takes a person to get to a retail center, the more likely he/she is to spend more time in the center and do more recreational activities. Besides, it was also found that the greater the variety of shops in a retail center, the more likely it is to attract customers who live farther away from the retail center, and they probably could achieve to their main goal for attracting more customers not only for purchasing but also for eating, browsing, and recreational activities. Also, the users who select these places consciously, and prefer to spend more time and do more recreational activities in these new modern units in comparison to public spaces, being in public spaces for users would be such necessary activities, not optional ones.

The results of the research that have been conducted in both case studies show even though retail centers were constructed to supply residents' needs, their social role is quite high due to the variety of facilities available, and the majority of customers attended the centers for reasons other than purchasing, the existence of recreational, service and welfare facilities can be the driving factors for attracting people. While retail centers were initially established as a modern and new phenomenon in cities, their patterns have changed across the world in the same way. Based on these findings, it appears that the primary motivation of retail centers in developed countries such as Spain is related to social activities, however, in developing countries, such as Iran, centers are increasing their social role to attract more visitors. Regarding the significance of this study and the rapid changes that have taken place in developing countries, further research is required since changes in developing countries are more influenced by imitation of patterns, rather than following a particular process and temporary policies. According to this study, the retail centers play a relatively strong social role besides their commercial role, and today, due to the widespread formation of retail centers in cities, strengthening the social role of these centers should be considered more than before. It was found that the diversity of retail centers can be one of the factors for promoting the social role of retail centers and also can affect the customers' travel time duration in the shopping center and since this research has been done on two retail centers that are located in two completely different countries in terms of culture and consumption pattern, but the results are almost similar for both of them, which shows the importance of promoting the social role of retail centers.

*6.1. Managerial implications*

Retail centers are dynamic business places that a notable number of customers are attracted to experience the pleasure of shopping, the results and discussion of this study provide significant practical notions and frameworks for the centers' managers to improve the quality of their centers also pay special attention to choosing the location of their retail centers. Since the results of this study proved that the time it takes for a customer to reach the retail center affects the time spent in the retail center, the managers of these shopping centers should consider this important factor in locating shopping centers. On the other hand, the confirmation of the second hypothesis of this research, which proves the effect of the diversity of shops in a shopping center on the time spent by a customer in that shopping center, sends a message to retail center managers that by increasing the diversity of brands and also, the variety of products sold in a mall can provide a guarantee of increasing the time spent by a customer in that mall.

*6.2 Limitation*

Like many other experimental researches, this one might also have some limitations in reference to sampling, collecting data, and generalizing the findings. Also, this study only has investigated two shopping centers with a similar character, on the other hand, the pattern of using the different types of transportation can also be one of the influential factors that can yield different results depending on the culture of using them in an individual country, so the specimen derived for the research probably not be enough to generalize the results of the other researches.

6.3 Future research prospects

The core idea of this research is to study the driving factors behind the social role of retail centers on recreational activities such as travel time and shop variety. There are very few studies available on the retail centers that have investigated the effect of time and shop variety simultaneous. Researchers exploring the area of retail centers are encouraged to accomplish comparative studies on different types of retail centers with different patterns of behavior and character, also there are many other factors that can be effective on the social role of retail centers as a place for doing recreational activities, like the price of the products, the sense of place and etc.

**Conflicts of Interest:** The authors declare no conflict of interest.